\documentclass[a4paper,11pt]{article}
\usepackage{pos}

\newcommand{\mA}{\mathcal{A}}

\newcommand{\mD}{\mathcal{D}}

\newcommand{\be}{\begin{equation}}
\newcommand{\ee}{\end{equation}}
\newcommand{\bea}{\begin{eqnarray}}
\newcommand{\eea}{\end{eqnarray}}
\newcommand{\ba}{\begin{align}}
\newcommand{\ea}{\end{align}}
\newcommand{\bse}{\begin{subequations}}
\newcommand{\ese}{\end{subequations}}
\newcommand{\comment}[1]{}

\title{Symplectic Quantization and Minkowskian Statistical Mechanics: simulations on a 1+1 lattice}
\ShortTitle{Symplectic Quantization}

\author*[a,b]{Martina Giachello}
\author[a,b]{Giacomo Gradenigo}
\author[c,d]{Francesco Scardino}

\affiliation[a]{Gran Sasso Science Institute, Viale F. Crispi 7, 67100 L'Aquila, Italy}
\affiliation[b]{INFN-Laboratori Nazionali del Gran Sasso, Via G. Acitelli 22, 67100 Assergi (AQ), Italy}
\affiliation[c]{Physics Department, INFN Roma1, Piazzale A. Moro 2, Roma, I-00185, Italy}
\affiliation[d]{Physics Department, Sapienza University, Piazzale A. Moro 2, Roma, I-00185, Italy}

\emailAdd{martina.giachello@gssi.it}

\abstract{
We introduce \textit{symplectic quantization}, a novel functional approach to quantum field theory which allows to sample quantum fields fluctuations directly in Minkowski space-time, at variance with {\it all} the traditional importance sampling protocols, well defined only for Euclidean Field Theory. This importance sampling procedure is realized by means of a deterministic dynamics generated by Hamilton-like equations evolving with respect to an auxiliary time parameter \( \tau \). In this framework, expectation values over quantum fluctuations are computed as dynamical averages along the trajectories parameterized by \( \tau \). Assuming ergodicity, this is equivalent to sample a microcanonical partition function. Then, by means of a large-M calculation, where M is the number of degrees of freedom on the lattice,  we show that the microcanonical correlation functions are equivalent to those generated by a Minkowskian canonical theory where quantum fields fluctuations are weighted by the factor $\exp(S/\hbar )$, with $S$ being the original relativistic action of the system \cite{Giachello:2024wqt}.}

\FullConference{%
 The 41st International Symposiumon Lattice Field Theory (LATTICE2024)\\
  28 July -3 August2024\\
  Liverpool, UK
}

\begin{document}
\maketitle

\section{Introduction}
\label{sec:introduction}
Since its introduction by Kenneth Wilson~\cite{W74}, lattice field theory has become a powerful tool for addressing non-perturbative problems in quantum field theory~\cite{C83,MM94}. The key limitation of present lattice field-theoretic methods is that they are well defined only for Euclidean theories. The rotation from real to immaginary time, inherent to all Euclidean theories, prevents in fact the study of all processes related to space-time causal structure, e.g., light-cone dynamics or scattering processes involving a different number of degrees of freedom in initial and final asymptotic states. These limitations motivate the need for an approach to quantum field theory allowing for the definition of a numerical protocol to sample quantum fluctuations directly in Minkowski space-time. A promising development in this direction is \textit{symplectic quantization}, introduced in~\cite{GL21,G21,GLS24,Giachello:2024wqt}. In this framework, a quantum field \( \phi(x,\tau) \), with \( x = (ct, \mathbf{x}) \), evolves with respect to an additional time parameter \( \tau \), which governs the sequence of quantum fluctuations on the discretized space-time lattice. Unlike Parisi-Wu stochastic quantization~\cite{PW81,DH87}, which also uses an auxiliary time variable but is limited to Euclidean space, the deterministic dynamics of symplectic quantization is well defined aso in Lorentzian space-time. 
\section{Symplectic Quantization}
\label{sec:SQ}
Symplectic quantization dynamics has been inspired in first instance by the stochastic quantization approach~\cite{PW81,DH87}. First of all one has to consider an additional time variable $\tau$ that parametrizes the dynamics of quantum fluctuations at each point of the space-time lattice, allowing the field $\phi(x,\tau)$ to depend on it. Let us remark that $\tau$ is distinct from the observer's time $x^0$, as discussed in detail in~\cite{Giachello:2024wqt,GL21,G21}. Symplectic quantization then exploits the existence of conjugated momenta related to the rate of variation of the field with respect to $\tau$:
\begin{equation}
\pi(x,\tau) \propto \dot{\phi}(x,\tau).
\end{equation}
Formally, one introduces the generalized Lagrangian:
\begin{equation}
\mathbb{L}[\phi,\dot{\phi}] = \int d^dx \left[ \frac{1}{2c_s^2} \dot{\phi}^2(x) + S[\phi] \right],
\end{equation}
where $c_s$ is a dimensionless constant (we will consider $c_s = 1$ for simplicity) and $S[\phi]$ is the standard relativistic action, which for a real scalar field with $\lambda\phi^4$ interaction reads as:
\begin{equation}
S[\phi] = \int d^dx \left( \frac{1}{2} \partial_\mu \phi(x) \partial^\mu \phi(x) - V[\phi] \right) \qquad\text{with}\qquad V[\phi] \frac{1}{2} m^2 \phi^2 - \frac{1}{4} \lambda \phi^4. 
\end{equation}
Performing a Legendre transform of the Lagrangian, we obtain the Hamiltonian:
\begin{equation}
\mathbb{H}[\phi, \pi] = \frac{1}{2} \int d^dx \, \pi^2(x) - S[\phi],
\end{equation}
which explicitly reads
\begin{equation}
\mathbb{H}[\phi, \pi] = \int d^dx \left[ \frac{1}{2} \pi^2(x) - \frac{1}{2} \left( \frac{\partial \phi}{\partial x^0} \right)^2 + \frac{1}{2} \sum_{i=1}^{d} \left( \frac{\partial \phi}{\partial x^i} \right)^2 + V[\phi] \right].
\end{equation}
One then assumes that the dynamics of quantum fluctuations is governed by the Hamilton equations:
\begin{equation}
\dot{\phi}(x) = \frac{\delta \mathbb{H}[\phi,\pi]}{\delta \pi(x)}, \quad \dot{\pi}(x) = - \frac{\delta \mathbb{H}[\phi,\pi]}{\delta \phi(x)},
\label{eq:ham-dym}
\end{equation}
which are equivalent to the following equation of motion:
\begin{equation}
\ddot{\phi}(x,\tau) = - \partial_0^2 \phi(x,\tau) + \sum_{i=1}^{d-1} \partial_i^2 \phi(x,\tau) - \frac{\delta V[\phi]}{\delta \phi(x,\tau)}.
\label{eq:eq-motion}
\end{equation}
At this stage the symplectic quantization approach is formally very similar to the microcanonical approach to quantum field theory proposed by Callaway and Raman \cite{Callaway:1983ee}, which is also based on a deterministic dynamics with respect to an additional time parameter $\tau$ and exploits the presence of additional conjugated momenta, but as all previous approaches is based on Euclidean Field Theory, i.e., the equation of motions for fields differ by a sign with respect to Eq.~\eqref{eq:eq-motion}:
\begin{equation}
\ddot{\phi}(x,\tau) =  \partial_0^2 \phi(x,\tau) + \sum_{i=1}^{d-1} \partial_i^2 \phi(x,\tau)  - \frac{\delta V[\phi]}{\delta \phi(x,\tau)}.
\label{eq:eq-motion-eucl}
\end{equation}
Let us now comment on how the symplectic quantization dynamics is related to the Feynman path-integral formulation of quantum field theory. Assuming the system samples the energy hypersurface uniformly in the long-time limit, the distribution of field configurations can be written as:
\begin{equation}
\rho_{\text{micro}}[\phi(x)] = \frac{1}{\Omega[\mA]} \delta\left( \mA - \mathbb{H}[\phi,\pi] \right),
\end{equation}
where $\Omega[\mA]$ is the microcanonical partition function:
\begin{equation}
\Omega[\mA] = \int \mathcal{D}\phi \, \mathcal{D}\pi \, \delta\left( \mA - \mathbb{H}[\phi,\pi] \right),
\end{equation}
where $\mathcal{D}\phi$ and $\mathcal{D}\pi$ represent standar measure for functional integration. From the microcanonical partition function one can then define a generalized microcanonical entropy:
\begin{equation}
\Sigma_{\text{sym}}[\mA] = \ln \Omega[\mA].
\end{equation}
By assuming a {\it bona-fide} ergodic hypothesis we claim the equivalence between time and ensemble averages of a generic fields observable $\mathcal{O}[\phi(x)]$:
\begin{equation}
\lim_{\Delta \tau \rightarrow \infty} \frac{1}{\Delta \tau} \int_{\tau_0}^{\tau_0 + \Delta \tau} d\tau \, \mathcal{O}[\phi(x,\tau)] = \int \mathcal{D}\phi \, \mathcal{D}\pi \, \rho_{\text{micro}}[\phi(x)] \, \mathcal{O}[\phi(x)].
\end{equation}
The Feynman path integral is then obtained by first Fourier transforming the partition function
\begin{equation}
\mathcal{Z}[z] = \int_{-\infty}^{\infty} dA \, e^{-izA} \, \Omega[A] = \int \mathcal{D}\phi \, \mathcal{D}\pi \, e^{- \frac{i}{2} z \int d^dx \, \pi^2(x) + iz S[\phi]},
\label{eq:FPI}
\end{equation}
then integrating over momenta and setting $z = \hbar^{-1}$:
\begin{equation}
\mathcal{Z}[\hbar] = \int_{-\infty}^{\infty} dA \, e^{-iA/\hbar} \, \Omega[A] \propto \int \mathcal{D}\phi \, e^{\frac{i}{\hbar} S[\phi]}.
\end{equation}
Let us notice that the need of Fourier rather than Laplace transform in Eq.~\eqref{eq:FPI}, which is the usual case for the change of ensemble in Statistical Mechanics, comes from the fact that, due to its relativistic nature, the action $S(\phi)$ cannot be made positive definite by adding a constant: it takes values in the whole real domain. This is why the Feynman path integral, differently from the canonical ensemble, is characterized by probability amplitudes rather than probabilities. 
\section{Hamiltonian dynamics on the lattice}
\label{sec:numerics}
\begin{figure}[h]
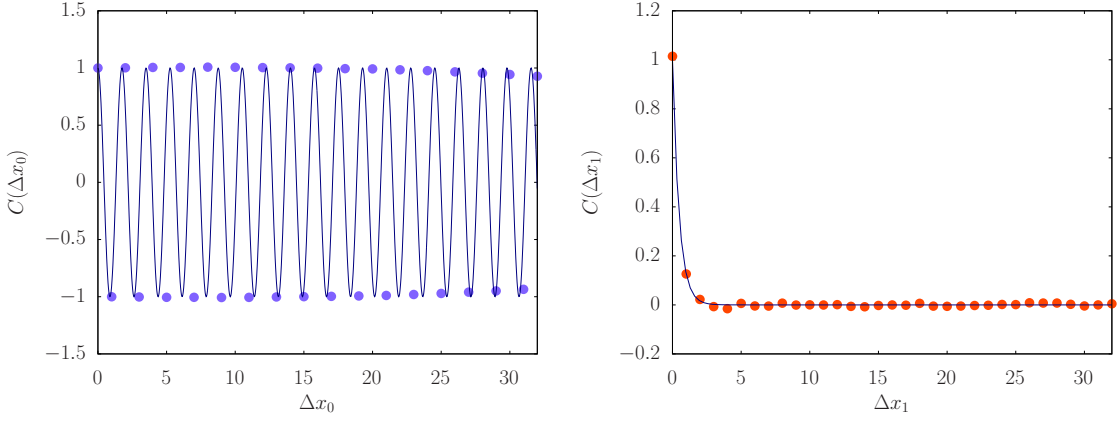

  \begin{minipage}{0.5\textwidth}
    \centering
    \includegraphics[width=\textwidth]{correlationfunction_t_fringe.png}
  \end{minipage}%
  \begin{minipage}{0.5\textwidth}
    \centering
    \includegraphics[width=\textwidth]{correlationfunction_x_fringe.png}
  \end{minipage}
    \caption{Two-point correlation function in real space for a $\lambda\phi^4$ theory in $1+1$ dimensions, lattice spacing $a=1.0$, lattice size $L=128$, mass $m=1.0$, and nonlinearity $\lambda=0.001$. \textbf{Right:} Exponential decay along the $x_1$ (spatial) axis. \textbf{Left:} Oscillations along the $x_0 = ct$ (temporal) axis.}
  \label{fig-4}
\end{figure}

As a first test of this approach, we performed numerical simulations on a 1+1-dimensional lattice. The discretized Hamiltonian theory is given by:
\begin{equation}
\mathbb{H}[\phi,\pi] = \frac{1}{2}\sum_{{\bf i} \in \mathbb{Z}^2} \left[ \pi_{\bf i}^2 + \frac{1}{a^2} \phi_{\bf i}\Delta^{(0)}\phi_{\bf i} - \frac{1}{a^2} \phi_{\bf i}\Delta^{(1)}\phi_{\bf i} + m^2\phi_{\bf i}^2 +\frac{\lambda}{4}\phi_{\bf i}^4 \right],
\end{equation}
where \(a\) denotes the lattice spacing, which has been fixed to \(a=1\) in our simulations, and \(\Delta^{(\mu)}\phi_{\bf i}\) represents the discrete one-dimensional Laplacian along the \(\mu\)-th coordinate axis:
\begin{equation}
\Delta^{(\mu)}\phi_{\bf i} = \phi_{{\bf i} + {\bf e}^\mu} + \phi_{{\bf i} - {\bf e}^\mu} - 2 \phi_{\bf i}.
\end{equation}
On the 1+1-dimensional lattice, the equations of motion are:
\begin{equation}\label{eq:eqofmotion}
\ddot{\phi}_{n,m}(t) = -\frac{\phi_{n+1,m} + \phi_{n-1,m} - 2 \phi_{n,m}}{a^2} + \frac{\phi_{n,m+1} + \phi_{n,m-1} - 2 \phi_{n,m}}{a^2} - m^2 \phi_{n,m} - \lambda \phi_{n,m}^3,
\end{equation}
where \(\phi_{n,m}(t)\) is the field at the lattice point with coordinates \((n,m)\). Since we deal with a classical dynamics we are faced with the choice of initial conditions, which must be consistent with having on average a quantum $\hbar$ of action for every degree of freedom, equally shared among the "positional", $\phi(k)$, and the "kinetic", $\pi(k)$, components. This "quantization" constraint can be fulfilled by assigning for each momentum $k$ in Fourier space the initial condition $|\pi(k_i;\tau=0)|^2 = \hbar$, which, assuming then equipartition, leads to the average $\langle \pi^*(k_i) \pi(k_i) \rangle = \frac{\hbar}{2}$. In our simulations we have fixed $\hbar=1$. This procedure is consistent with assuming that the scale $\hbar$ plays the role of temperature in the generalized microcanonical ensemble of symplectic quantization:
\begin{align}
    \frac{1}{\hbar} = \frac{d\Sigma_{\text{sym}}(\mathcal{A})}{d\mathcal{A}}.
\end{align}
The above strategy for setting the scale of the action per degree of freedom is analogous to the one usually considered to fix the temperature scale in molecular dynamics simulations. 
\\
Our main result is that, by simulating the dynamics of Eq.~(\ref{eq:eqofmotion}) for small values of the non-linearity coefficient $\lambda$, we are able to recover the qualitative form of the Feynman propagator for a scalar field theory simulated on an 1+1-dimensional lattice with side $L=128$. By setting $\lambda = 0.001$, we run the deterministic dynamics of Eq.~\eqref{eq:ham-dym} until the system reaches a stationary state, say at time $\tau_{\text{eq}}$. We then study the two-point correlation function $G(x,y) = \langle \phi(x) \phi(y)\rangle$ by computing the ensemble average as a sliding window average along the dynamics for times larger than the equilibration time, $\tau > \tau_{\text{eq}}$:
\begin{align}
\langle \phi(x) \phi(y) \rangle = \frac{1}{\Delta \tau} \sum_{i=0}^M \phi(x,\tau_{\text{eq}} + \tau_i) \phi(y, \tau_{\text{eq}} + \tau_i),
\end{align}
where $\tau_i = i\cdot\delta\tau$ and $\Delta \tau = M \delta\tau$. By doing this we recover the causal structure of the real-space correlation function corresponding to the Feyman propagator, i.e., with undamped oscillations along the time-like direction, $\Delta_F (x-y)\sim e^{im |x-y|}$,  and exponential decay along the space-like one, $\Delta_F (x-y)\sim e^{-m |x-y|}$. These two behavious are reported in Fig.~\ref{fig-4}. The only limitation of our numerical study is that, so far, the value of the mass extracted from two-point correlation functions, $m = 2.06 \pm 0.04$, differs from the input mass $m=1.0$, which calls for a more detailed study of finite-size effects. For a full account of the numerical simulations method, including the definition of the particular {\it "fringe"} boundary conditions needed to see the undamped propagation of signals along the time-like directon, we refer the reader to \cite{Giachello:2024wqt}.
\section{Canonical form of Minkowskian statistical mechanics}
\label{sec:pertequiv}
In the previous section we have shown that the microcanonical dynamics of a theory with additional momenta allows to recover qualitatively well the shape of the free Feynman propagator for a small value of the interaction constant $\lambda$.  It is therefore legitimate to wonder which is the precise relation between the correlation functions obtained in this generalized microncanonical ensemble and the one generated by the Feynman path integral and/or the corresponding Euclidean Field Theory. As a first step in this direction we propose an explicit calculation of the microcanonical partition function in the large-$M$ limit, where $M$ is the number of degrees of freedom. The calculation shows that also for this peculiar system the sampling of the microcanonical ensemble is formally equivalent to the sampling of a "canonical" one at "temperature" $\hbar$  and with effective energy precisely corresponding to the relativistic action $S(\phi)$, i.e., fields fluctuations are sampled with probability $\exp(S/\hbar)$.
The explicit computation of the microcanonical partition function in the large-$M$ limit proceeds then as follows. As is customary for the purpose of computing correlation functions, we assume the presence of an external source $J(x)$ linearly coupled to the field:
\begin{align}
  \Omega[\mA,J] = \int \mD\phi\mD\pi~\delta\left(\mA-\mathbb{H}[\phi,\pi]+\int d^dx\,J(x) \phi(x)\right).
  \label{eq:gen-micro}
\end{align}
First we assume an expansion of the field in an orthonormal basis as follows~\cite{STROMINGER}:
\begin{equation}\label{eq:expansion}
	\phi(x) = \sum_{n=1}^M \phi_n(x) c_n \quad \text{with} \quad \int d^dx\, \phi_n(x) \phi_m(x) = \delta_{mn}.
\end{equation}
The finite measure over field configurations becomes:
\begin{equation}
	\int \mathcal{D}_M \phi \equiv \prod_{n=1}^M \int_{-\infty}^{\infty} dc_n.
\end{equation}
In a $d$-dimensional box with volume $L^d$ and lattice spacing $a$, the number of basis functions is:
\begin{equation}\label{eq:Mlambda}
	M = \frac{L^d}{a^d} = \frac{1}{\pi^d} L^d \Lambda^d,
\end{equation}
where $\Lambda = \pi/a$ is the momentum cutoff. We denote the limit $M \to \infty$ as the {\it field} limit, which can be reached either as the thermodynamic limit ($a$ fixed, $L \to \infty$) or as the continuum limit ($L$ fixed, $a \to 0$).  By lightening the notation according to the following conventions
\begin{align}
        \pi^2 \equiv  \int d^dx\, \pi^2(x)~~~~~~~~~~~~~~
        J \cdot \phi \equiv \int d^dx\, J(x) \phi(x).
\end{align}
we can then rewrite the partition function on the lattice as:
\begin{equation}
	\Omega[\mA,J] = \int \mathcal{D}\phi_M\mathcal{D}\pi_M\, \delta\left(\mA-\frac{\pi^2}{2} +S[\phi]+J\phi\right)\,.
\end{equation}
The integration over $\pi$ can be done by taking advantage of the following formula, valid for $R^2>0$:
\begin{align}
  I_M(R) = \int_{-\infty}^\infty dx_1\ldots dx_M ~\delta\left( \frac{1}{2}\sum_{i=1}^M x_i^2 - R^2 \right) = \frac{(2\pi)^{\frac{M}{2}}}{\Gamma\left(\frac{M}{2}\right)}~R^{M-2}, 
\end{align}
from which, putting $R = (\mA+S[\phi]+J\phi)^{\frac{1}{2}}$, we get:
\begin{align}
  \label{eq:omega-step}
  \Omega[\mA,J] = \frac{(2\pi)^{\frac{M}{2}}}{\Gamma\left(\frac{M}{2}\right)} \int \mathcal{D}\phi_M\left(\mA+S[\phi]+J\phi\right)^{\frac{M}{2}-1}.
\end{align}
The positivity of $R^2=\mA+S[\phi]+J\phi$, which is crucial for the whole calculation, is ensured by construction of the microcanonical ensemble, since the kinetic energy obtained from conjugate momenta $\frac{1}{2}\pi^2$ is positive definite. In order to consider a large-$M$ limit in the computation is convenient to rewrite the partition function in Eq.~\eqref{eq:omega-step} as
\begin{align}
	\Omega[\mA,J] = \kappa_M~\int \mathcal{D}\phi_M\,e^{(\frac{M}{2}-1)\ln\left(\mA+S[\phi]+J\phi\right)},
    \label{eq:Omega-less-momenta}
\end{align}
where $\kappa_M= (2\pi)^{\frac{M}{2}}/\Gamma\left(M/2\right)$. In order to fulfill the same quantization constraint used for numerical simulation we assign also now $\hbar$ to every degree of freedom, equally sharing this amount among "positional" and "kinetic" components. Since momenta have already been integrated out, in expression Eq.~\eqref{eq:Omega-less-momenta} we need to fix $\mA$ to half of the total value, since we need to account only for "positional" degrees of freedom, namely we write
\begin{align}
\mA_z = \frac{\hbar M}{2z} 
\end{align}
We have introduced at this point the dimensionless parameter $z$ in order to be able to tune the value of the average "action quantum" per degree of freedom in the final expression and also to highlight how the present theory connects to ordinary Feynman path integral by analytic continuation in $z$. We now proceed to expand the partition function $\Omega[\mA_z,J]$: in powers of $J$ so that we can write explicitly the generating functional in terms of correlators. Then, by ignoring the subleading $O(1)$ in $M$ term in the exponent, we get:
\begin{align}
	\Omega[\mA_\hbar,J] &= ~\kappa_M \left(\frac{\hbar M}{2z}\right)^{\frac{M}{2}} \sum_{n=0}^{\infty}\frac{1}{n!}\left(\frac{z}{\hbar}\right)^n\left(\frac{2}{M}\right)^n\frac{\Gamma(\frac{M}{2}+1)}{\Gamma(\frac{M}{2}+1-n)}\nonumber\\
	&\times\int d^dx_1\ldots d^dx_n~J(x_1)\ldots J(x_n) \int \mathcal{D}_M\phi~\phi(x_1)\ldots \phi(x_n) \left(1+\frac{2}{M}\frac{z}{\hbar}S[\phi]\right)^{\frac{M}{2}-n}\,.
\end{align}
Now, we proceed to expand $\left(1+\frac{2}{M}\frac{z}{\hbar}S[\phi]\right)^{\frac{M}{2}-n}$ around $1/M$. A tedious but straightforward calculation yields:
\begin{equation}
  \left(1+\frac{2}{M}\frac{z}{\hbar}S[\phi]\right)^{\frac{M}{2}-n} =
  e^{\frac{z}{\hbar}S[\phi]}\nonumber \exp\Bigg(\sum_{j=1}^{\infty}(-1)^j~\frac{2(2j-2)!!}{(j+1)!}\left(\frac{z}{\hbar}\frac{S[\phi]}{M}\right)^j~[j S[\phi]+(j+1)n]\Bigg)\,.
\end{equation}
from which we have
\begin{align*}
	\Omega&[\mA_z,J] = \kappa_M~\left(\frac{\hbar M}{2z}\right)^{\frac{M}{2}}~\sum_{n=0}^{\infty}\frac{1}{n!}\left(\frac{z}{\hbar}\right)^n\left(\frac{2}{M}\right)^n\frac{\Gamma(\frac{M}{2}+1)}{\Gamma(\frac{M}{2}+1-n)} \nonumber\\
	&\times\int d^dx_1\ldots d^dx_n~J(x_1)\ldots J(x_n)~\left\langle \phi(x_1)\ldots \phi(x_n)~e^{\sum_{j=1}^{\infty}(-1)^j~\frac{2(2j-2)!!}{(j+1)!}\left(\frac{z}{\hbar}\frac{S[\phi]}{M}\right)^j~[j S[\phi]+(j+1)n]}\right\rangle,
\end{align*}
where the expectation value, denoted $\langle\,\cdot\,\rangle$, is taken with respect to $\int \mathcal{D}_M\phi\,e^{\frac{z}{\hbar}S[\phi]}$ with $S[\phi]$ being the renormalized action. Consider now the correlators:
\begin{align}
  &\left\langle \phi(x_1)\ldots \phi(x_n)~e^{\sum_{j=1}^{\infty}(-1)^j~\frac{2(2j-2)!!}{(j+1)!}\left(\frac{z}{\hbar}\frac{S[\phi]}{M}\right)^j~[j S[\phi]+(j+1)n]}\right\rangle \nonumber \\
  &=\left\langle \phi(x_1)\ldots \phi(x_n) \right\rangle+\sum_{j=1}^{\infty} \frac{c_j(M,n)}{M^j} \left\langle \phi_1(x_1)\ldots\phi_n(x_n)~S^{j}[\phi]\right\rangle,
  \label{eq:latticeomega2}
\end{align}
where in the second line we performed a large-$M$ expansion of the exponential: It turns out that the coefficients $c_j(M,n)$ are polynomials in $M$ with an asymptotic behavior of the kind $\frac{c_j(M,n)}{M^j} = o\left(\frac{1}{M}\right)$. We need now to ascertain whether $\Omega[\mA_z,J]$ has a sensible field limit, $M\rightarrow\infty$. We begin by noticing that $\lim_{M\to\infty}\left(2/M\right)^n \Gamma(\frac{M}{2}+1)/\Gamma(\frac{M}{2}+1-n)\to 1$, which tells us that the coefficient of each term in the sum goes to unity in the large-$M$ limit. It is at this point very reasonable assumption that all the insertions of powers of the renormalized action $c_j(M,n) S[\phi]^{j}/M^j$ in the correlators appearing in Eq.~\eqref{eq:latticeomega2} go smoothly to zero in the field limit: this is for instance equivalent to say that in the continuum limit the renormalized action remains finite in a finite volume, i.e., $S[\phi]/M \rightarrow 0$ when $M\rightarrow\infty$. Clearly at this step of the caculation we made an assumption which does not equally apply to both the continuum and thermodynamic limit, since in the thermodynamic limit we obviously have $S[\phi] \sim M$. Therefore since now on we will only speak about the continuum limit, where we have:
\begin{equation}
\lim_{M\to\infty}\left\langle \phi_1(x_1)\ldots,\phi_n(x_n) \frac{c_j(M,n)}{M^j}S^{j}[\phi]\right\rangle = 0 ~~~~~~~~~\forall ~ j,
\label{eq:null-insertion}
\end{equation}
which finally leads to:
\begin{align}
\Omega[\mA_z,J] &= \kappa_M~\left(\frac{ \hbar M}{2z}\right)^{\frac{M}{2}}~\sum_{n=0}^{\infty}\frac{1}{n!}\left(\frac{z}{\hbar}\right)^n
~\int d^dx_1\ldots d^dx_n~J(x_1)\ldots J(x_n)~\left\langle \phi(x_1)\ldots \phi(x_n)\right\rangle  \nonumber \\
& = \kappa_M~\left(\frac{ \hbar M}{2z}\right)^{\frac{M}{2}} \int \mathcal{D}\phi_M~e^{\frac{z}{\hbar}S[\phi]}~\sum_{n=0}^{\infty}\frac{1}{n!}\left(\frac{z}{\hbar}\right)^n\times\\
&\qquad\qquad\times\int d^dx_1\ldots d^dx_n~J(x_1)\ldots J(x_n)~\phi(x_1)\ldots \phi(x_n) \nonumber \\
& = \kappa_M~\left(\frac{ \hbar M}{2z}\right)^{\frac{M}{2}} \int \mathcal{D}\phi_M~e^{\frac{z}{\hbar}S[\phi]+\frac{z}{\hbar} \int d^{d}x~J(x)\phi(x)}\,.
\label{eq:ultimate-result}
\end{align}
From Eq.~\eqref{eq:ultimate-result} we can therefore conclude that, up to irrelevant multiplicative constants, the symplectic quantization microcanonical generating functional in then {\it continuum} limit takes the form:
\begin{align}
	\Omega[\hbar/z,J]=\int \mathcal{D}\phi\,e^{\frac{z}{\hbar}S[\phi]+\frac{z}{\hbar}J\phi}.
    \label{eq:final-Omega}
\end{align}
The choice of $z$ corresponding to the simulations presented in the first part of this work is $z=1$: in this case the expression of $\Omega[\hbar,J]$ obtained in Eq.~\eqref{eq:final-Omega} tells us that the correlation function measured in microcanonical dynamical simulations must be identical, to the leading order in $M$, to those obtained from a canonical probability distribution 
\begin{align}
P[\phi] = e^{S[\phi]/\hbar}/\Omega[\hbar].
\label{eq:final-P}
\end{align}
Let us conclude with two main remarks about the results in Eqns.~\eqref{eq:final-Omega}, \eqref{eq:final-P}. First of all we have shown that the microcanonical sampling is equivalent to the sampling from a probability distribution $P[\phi]$ which is well defined for an interacting theory with a potential bounded from below, since for configuration of the field with large values and smooth variations we have approximatively
\begin{align}
    e^{S[\phi]/\hbar} \sim e^{-V[\phi]/\hbar}.
\end{align}
This is completely in agreement with the results of numerical simulations in the microcanonical ensemble, where the Hamiltonian dynamics of the free theory develops run-away solutions. Second, the result of our derivation in Eq.~\eqref{eq:final-Omega} shows us that this new "canonical Minkowskian measure" can be connected to the standard Feynman path integral by means of analytic continuation in the dimensionless parameter $z$. More investigations in this direction are actually in progress.  

\newpage
\bibliographystyle{JHEP}
\bibliography{reference-1}

\end{document}